\documentclass{aastex62}

\usepackage{multirow}

\graphicspath{{./}{figures/}}

\received{}
\revised{}
\accepted{}
\submitjournal{ApJ}

\shorttitle{3D statistical kinematics of CMEs}
\shortauthors{Majumdar et al.}
\begin{document}

\title{An Insight into the Coupling of CME Kinematics in Inner and Outer Corona and the Imprint of Source Regions}

\correspondingauthor{Dipankar Banerjee}
\email{dipu@iiap.res.in, dipu@aries.res.in}

\author[0000-0002-6553-3807]{Satabdwa Majumdar}
\affiliation{Indian Institute of Astrophysics,2nd Block, Koramangala, Bangalore, 560034, India}
\affiliation{Pondicherry University, Chinna Kalapet, Kalapet, Puducherry 605014}

\author[0000-0001-8504-2725]{Ritesh Patel}
\affiliation{Indian Institute of Astrophysics,2nd Block, Koramangala, Bangalore, 560034, India}
\affiliation{Aryabhatta Research Institute of Observational Sciences, Beluwakhan, 263001, Uttarakhand}
\affiliation{ University of Calcutta, 87, 1, College St, College Square, Kolkata, West Bengal 700073}

%\nocollaboration

\author[0000-0002-6954-2276]{Vaibhav Pant}
\affiliation{Aryabhatta Research Institute of Observational Sciences, Beluwakhan, 263001, Uttarakhand}
%\affiliation{Institute de Astroficisia de Canarias (IAC), Tenerife}
%\collaboration{(AAS Journals Data Scientists collaboration)}

\author[0000-0003-4653-6823]{Dipankar Banerjee}
\affiliation{Indian Institute of Astrophysics,2nd Block, Koramangala, Bangalore, 560034, India}
\affiliation{Aryabhatta Research Institute of Observational Sciences, Beluwakhan, 263001, Uttarakhand}
\affiliation{Center of Excellence in Space Science, IISER Kolkata, Kolkata 741246, India}
%\collaboration{(LaTeX collaboration)}



\begin{abstract}

Despite studying Coronal Mass Ejections (CMEs) for several years, we are yet to have a complete understanding of their kinematics. In this regard, the change in kinematics of the CMEs, as they travel from the inner corona ($<$ 3R$_\odot)$ to the higher heights is essential. We do a follow up statistical study of several 3D kinematic parameters of 59 CMEs studied by \cite{2020ApJ...899....6M}. The source regions of these CMEs are identified and classified as Active Regions (ARs), Active Prominences (APs), and Prominence Eruptions (PEs). We study several statistical correlations between different kinematic parameters of the CMEs. We show that the average kinematic parameters change as they propagate from the inner to the outer corona, indicating the importance of the region where normally the common practice is to perform averaging. We also find that the parameters in the outer corona is highly influenced by those in the inner corona, thus indicating the importance of inner corona in the understanding of the kinematics. We further find that the source regions of the CMEs tend to have a distinct imprint on the statistical correlations between different kinematic parameters, and that an overall correlation tends to wash away this crucial information. The results of this work lends support towards possibly different dynamical classes for the CMEs from active regions and prominences which is manifested in their kinematics. 
%A detailed study on the ejection mechanism for CMEs connected to these different sources will help in better establishing our claim.

\end{abstract}



\section{Introduction} \label{sec:intro}

Coronal mass ejections (CMEs) are large scale eruption of plasma and magnetic field from the solar corona into the heliosphere \citep{article}. Their speed ranges from a few hundreds to a few thousands of km s$^{-1}$, and acceleration ranging from few tens to a few 10$^4$ m s$^{-2}$ \citep[for a review, see][]{article}. CMEs are also the major drivers of space weather, as they are capable of producing shock waves, interplanetary disturbances, causing huge technological damages \citep{1993JGR....9818937G}. Thus, they are of interest from both scientific and technological point of views, and hence a good understanding of their kinematics is essential. It is understood that the kinematics of CMEs are governed by the interplay of three forces, namely the Lorentz force, the gravitational force and the viscous drag force  \citep{Wood1999ApJW, Zhang2001, article,2007SoPh..241...85V}. As a result of these forces, CMEs follow a three phase kinematic profile. According to \cite{Zhang_2006}, the first, initial phase is a slow rise phase, followed by an impulsive acceleration phase (observed as rapid increase in their velocity) and then the final phase where the CMEs propagate with little or no acceleration. In this regard, the first two phases are usually over in the low coronal heights ($< 3R_{\odot}$) \citep{Temmer_2008, 2011ApJ...738..191B, ciisco2020}. At later stages of their evolution, CMEs experience drag due to solar wind resulting in the deceleration \citep{nat2000,Moon2002ApJM,article}. So as kinematics of CMEs change from inner to outer corona, averaging of the different kinematic parameters over their entire trajectory might lead to washing away of a lot of crucial information that might hold clue to the coupling of kinematics of CMEs in the inner corona, to the heliosphere.

CMEs are also known to be associated with different source regions, Active Regions (ARs) and Prominence Eruptions (PEs) \citep{Subramanian_2001,Moon2002ApJM,2020ApJ...899....6M}. CMEs associated with ARs are known to be mostly impulsive whereas the ones associated with PEs are gradual CMEs \citep{MacQueen1983SoPhM,Sheeley1999JGR}. Recently \cite{pant_2021} reported the influence of different source regions on the width distribution of CMEs. Now, whether these different source regions have any clear imprint on different kinematic properties of CMEs is still not known clearly. 
%A knowledge of the location of the source region of the CMEs in terms of the latitude and longitude provides a correction factor to get the actual quantities from the projected quantities.

A major concern in the study of CME kinematics is regarding the measurements that are carried out in the plane of the sky, thus leading to projection effects in the measured quantities \citep{Balmaceda2018ApJ}. A primary step to minimise such projection effects, is to connect the CMEs to their source regions on the disk of the Sun. An even better way to remove the projection effects is to use 3D reconstruction techniques. In this regard, several works based on the tracking of CMEs in 3D have been reported \citep{Thernisien2006ApJ,Mierla2008SoPh, 2009SoPh..256..111T,Moran2010ApJ,Joshi2011ApJ,Sarkar2019ApJ}. A method based on forward modelling to fit the CME flux rope on multi-vantage-point images was also developed assuming the self-similar expansion of CMEs \citep{Thernisien2006ApJ,2009SoPh..256..111T}, which was termed as the Graduated Cylindrical Shell (GCS) model. Thus a study based on the fitted parameters of the model will be free from projection effects. Recently, \cite{2020ApJ...899....6M} connected 3D profiles of width evolution and acceleration to report on the observational evidence of the imprint of the height of influence of Lorentz force on the 3D kinematics. One of the most significant relevance of these 3D parameters is in the context of arrival time prediction of CMEs. Several models have been developed that takes the average speed of the CME as input to predict their arrival times \citep[for a review see][]{2014SpWea..12..448Z,2018SpWea..16.1245R}.

Since the kinematics of the CMEs change as they propagate outwards, it is important to look at the coupling of the kinematics of the CMEs from the inner to the outer corona. As it is already reported several times in the past that quiescent prominences and active regions tend to classify the ejected CMEs from them into two dynamical classes, with the former tending to be gradual CMEs while the later being impulsive CMEs \citep{Sheeley1999JGR}, it is worth looking at the manifestation of this distinction in the behaviour of different kinematic parameters that reflect the kinematics of a CME. In this work, keeping in mind the above existing shortcomings in our understanding of CME kinematics, we try look into the correlation between different 3D kinematic parameters of the CMEs as they evolve from the inner to the outer corona. With the additional information of the source region of the CMEs, we also look into the imprint of the source regions (if any) on the behaviour of these different 3D kinematic parameters. We study the same 59 events as studied by \cite{2020ApJ...899....6M} and follow the same analysis. In this context, it should be noted that a shock spheroid model is also available as a part of the GCS model (as reported earlier by \cite{hess_zang_2014}) for fitting the shock front ahead of the flux-rope. Since we were not interested in the shock dynamics and our aim was focused on the CME kinematics, we used only the flux-rope GCS model in our work. Alike \cite{2020ApJ...899....6M}, two vantage point observations are used for fitting the GCS model. It must be noted that provision for the use of a third vantage point in the form of observations from  Large Angle and Spectrometric Coronagraph \citep[LASCO;][]{Brueckner95} on-board the Solar and Heliospheric Observatory (SOHO) can be used for better constraining parameters like the tilt-angle \citep[see][]{2009SoPh..256..111T}. Since the aim of this work was to study the radial kinematics of CMEs, hence the tilt-angle parameter was not used in our analysis, thus reducing the need for the third vantage point. Also, the field of view of LASCO starts well beyond the starting FOV of COR-1, and since we don’t want to include Extreme Ultraviolet observations with white light, in order to keep consistency, we did not include LASCO observations for the GCS fitting.  In section \ref{sec2} we outline the data source used and the working method, followed by our results in section \ref{results} and we summarize the main conclusions from our work in section \ref{conclusion}

\section{Data and Method} \label{sec2}
\subsection{Data Source and Data Preparation}
The data used for this work is primarily taken from COR-1 and COR-2 on-board Sun Earth Connection Coronal and Heliospheric Investigation (SECCHI) package \citep{secchi} of the twin spacecraft Solar Terrestrial Relations Observatory \citep[STEREO;][]{2008SSRv..136....5K}. Also data from different passbands of Atmospheric Imaging Assembly (AIA) on-board Solar Dynamics Observatory \citep[SDO;][]{aia} and Extreme ultraviolet Imaging Telescope \citep[EIT;][]{SOHOEIT} on-board SOHO were used to identify the source regions of CMEs that were coming from the front side of the Sun. For more details on the data source, please refer \cite{2020ApJ...899....6M}. 

\subsection{Parameter description}
\label{sec:param}
% In this work, we aim to understand how different kinematic parameters of a CME are related to each other. 
To understand the relationship between the different parameters associated with kinematics of CME as the CME propagates from the inner to the outer corona, we list out the different parameters used in this work in Table \ref{table1} and define them as follows. 

\begin{itemize}
    \item a$_{\mathrm{max}}\;($V$_{\mathrm{max}})$ - peak acceleration (velocity) of the CME in the entire (COR-1 and COR-2) field of view (FOV)
    %\item $V_{max}$ - peak velocity of the CME in the entire FOV
    \item V$_{\mathrm{amax}}$ - Velocity of the CME at a$_{\mathrm{max}}$
    \item V$_{\mathrm{lin}}$ - average velocity of the CME from linear fit to the height-time data for the entire FOV
    \item V$_{\mathrm{mi}}\;($a$_{\mathrm{mi}})$ - mean velocity (acceleration) in the inner corona ($<\,3\,$R$_{\odot}$) computed by taking the mean of the velocity (acceleration)-time data points obtained from derivatives of the height-time data 
    \item a$_\mathrm{m}\;($V$_\mathrm{m})$ - Overall mean acceleration (velocity) in the entire FOV by computing the mean of the acceleration (velocity)-time data points obtained from derivatives of the height-time data 
   % \item $a_{mi}$ - mean acceleration in the inner corona ($<\;3\,$R$_{\odot}$) by taking mean of the acceleration-time data points obtained from spline fit to height-time data
  %  \item $a_m$ - mean acceleration in the entire FOV by taking mean of the acceleration time data points obtained from spline fit to height time data
    \item a$_{\mathrm{const}}$ - constant acceleration in the entire FOV, found from quadratic fit to the height-time data for the entire FOV. 
    \item h$_{\mathrm{amax}}\;($h$_{\mathrm{vmax}})$ - height at which peak acceleration (velocity) was attained
  %  \item $h_{amax}$ - height at which peak acceleration ($a_{max}$) was attained
    
\end{itemize}

\begin{center}
\begin{table}[]
    \centering
    \begin{tabular}{c|c|c|c|c}
    \hline \hline
    Parameters studied & Correl. Coeff. (CC) & Crit. Correl. Coeff. (CCC) & P-values & Empirical relation  \\
    \hline \hline
      \multirow{5}{*}{$(a)\:$a$_{\mathrm{max}}$ v/s V$_{\mathrm{max}}$}  & 0.63 (overall) & 0.25 & $7.8\:\times\:10^{-8}$ &  $ \mathrm{a}_{\mathrm{max}} = 10^{-0.35}\, \mathrm{V}_{\mathrm{max}}^{1.21}$\\
         & 0.45 (AP) & 0.46 & 0.053 & -- \\
         & 0.91 (AR) & 0.44 & $4.3\:\times\:10^{-8}$ &  $\mathrm{a}_{\mathrm{max}} = 10^{-2.71}\, \mathrm{V}_{\mathrm{max}}^{1.98}$\\
         & 0.41 (PE) & 0.44 & 0.075 & -- \\
        \hline
        \multirow{5}{*}{$(b)\:$V$_{\mathrm{amax}}$ v/s a$_{\mathrm{max}}$}  & 0.57 (overall) & 0.25 & $3.2\:\times\:10^{-6}$ & $\mathrm{V}_{\mathrm{amax}} = 10^{1.30}\, \mathrm{a}_{\mathrm{max}}^{0.43}$\\
          & 0.42 (AP) & 0.46 & 0.07 & -- \\
          & 0.77 (AR) & 0.44 & $7.4\:\times\:10^{-5}$ & $\mathrm{V}_{\mathrm{amax}} = 10^{1.31}\, \mathrm{a}_{\mathrm{max}}^{0.44}$\\
          & 0.57 (PE) & 0.44 & 0.009 & -- \\
           \hline
          \multirow{5}{*}{$(c)\:$V$_{\mathrm{lin}}$ v/s V$_{\mathrm{mi}}$}  & 0.68 (overall) & 0.25 & $2.6\:\times\:10^{-9}$ & $ \mathrm{V}_{\mathrm{lin}} = 10^{1.51}\, \mathrm{V}_{\mathrm{mi}}^{0.46}.$\\
            & 0.78 (AP) & 0.46 & $9.8\:\times\:10^{-5}$ & --\\
            & 0.63 (AR) & 0.44 & 0.003 & --\\
            & 0.78 (PE) & 0.44 & $5.8\:\times\:10^{-5}$ & --\\
           \hline
          \multirow{5}{*}{$(d)\:$V$_{\mathrm{max}}$ v/s V$_\mathrm{m}$}  & 0.73 (overall) & 0.25 & $4.6\:\times\:10^{-11}$ & $ \mathrm{V}_{\mathrm{max}} = 10^{1.05}\, \mathrm{V}_{\mathrm{m}}^{0.72}$\\
           & 0.88 (AP) & 0.46 & $5.1\:\times\:10^{-7}$ & --\\
           & 0.67 (AR) & 0.44 & 0.001 & --\\
           & 0.66 (PE) & 0.44 & 0.002 & --\\
           \hline
          
          \multirow{5}{*}{$(e)\:$a$_{\mathrm{const}}$ v/s V$_{\mathrm{mi}}$}  & -0.67 (overall) & -0.25 & $8.9\:\times\:10^{-9}$ & $\mathrm{a}_{\mathrm{const}} = 545.6\, \mathrm{V}_{\mathrm{mi}}^{-0.52}$\\
           & -0.62 (AP) & -0.46 & 0.006 & --\\
           & -0.61 (AR) & -0.44 & 0.005 & --\\
           & -0.30 (PE) & -0.44 & 0.202 & --\\
           \hline \hline
    \end{tabular}
    \caption{Summary of the fitted parameters with the Correlation Coefficients (CCs),Critical Correlation Coefficients (CCCs), p-values and the fitted relations. Note that apart from the overall empirical relation, the same is also shown separately only for the source regions for which the CC is distinctly different and higher than the CC of others. In ($a$) and ($b$) the CC for AR is much higher than others. In ($c$) and ($d$), the individual CCs are similar, while in ($e$) the CC for AR and AP are similar to the overall CC, and the CC for PE is much poor.}
    \label{table1}
\end{table}
\end{center}

%\multirow{4}{*}{$(e)\:a_{max}$ v/s $V_{mi}$}  & 0.33 (overall) & $a_{max} = 10^{2.01}\, V_{mi}^{0.49}$\\
           %& 0.60 (AP) & $a_{max} = 10^{1.42}\, V_{mi}^{0.73}$\\
           %& 0.14 (AR) & --\\
           %& 0.37 (PE) & --\\
           %\hline

\section{Results\label{results}}

We derived the parameters mentioned in Section \ref{sec:param} from the velocity and acceleration calculations of the 59 CMEs described in \cite{2020ApJ...899....6M}, studied their correlations and established several empirical relations amongst them. In this regard, some of the CMEs studied in \cite{2020ApJ...899....6M} show deceleration very close to the Sun \citep[see Figure 2 of][]{2020ApJ...899....6M}. It should be noted that we have used coronagraph images with a FOV starting from 1.4 R$_{\odot}$ (in the plane of sky) and in some cases the CMEs have already reached 2 R$_{\odot}$ (projected height) for the first measurement. According to \cite{2011SoPh..271..111G}, the impulsiveness of the CMEs occurs below 1.5 R$_{\odot}$. As the above mentioned CMEs are well beyond this height for the initial points of measurement, it is possible to have deceleration at the start of the coronagraph FOV. Further, in this regard, \cite{2017SoPh..292..118S} had reported that for fast CMEs, solar wind drag can act earlier, leading to the deceleration. It should also be noted that the widths mentioned in Table 1 of  \cite{2020ApJ...899....6M} is the value of the half-angle parameter measured at a particular instant of time and height (as mentioned in the table), beyond which the parameter experiences further evolution. Further, while comparing the GCS width with the width given in the Coordinated Data Analysis Workshop (CDAW) catalogue \citep[][]{Gopalswamy2009EM&PG}, it must be kept in mind that the complete width of the CME as measured from the GCS model is estimated as $2(half\,angle + Sin^{-1}(aspect\, ratio))$, while CDAW provides the final projected width, and for halo or partial-halo CMEs, the projected width becomes an over-estimate. In Table~\ref{table1} of this work, we provide the Pearson's correlation coefficient (CC) and for better appreciation of our results, the Pearson's critical correlation coefficients (CCCs). We also provide the associated p-value which shows the statistical significance of the correlation. The average significance level ($\alpha$) for the correlations was found to be $0.05$ on average, thus correlations with p-values lesser than $0.05$ and CCs higher than CCCs implies statistically significant result. In Figure \ref{fig1}(a) We plot a$_{\mathrm{max}}$ versus V$_{\mathrm{max}}$. We find that the two parameters are positively correlated with a CC of $0.63$ and can be described by the following relation:

\begin{figure*}[h]
\gridline{\fig{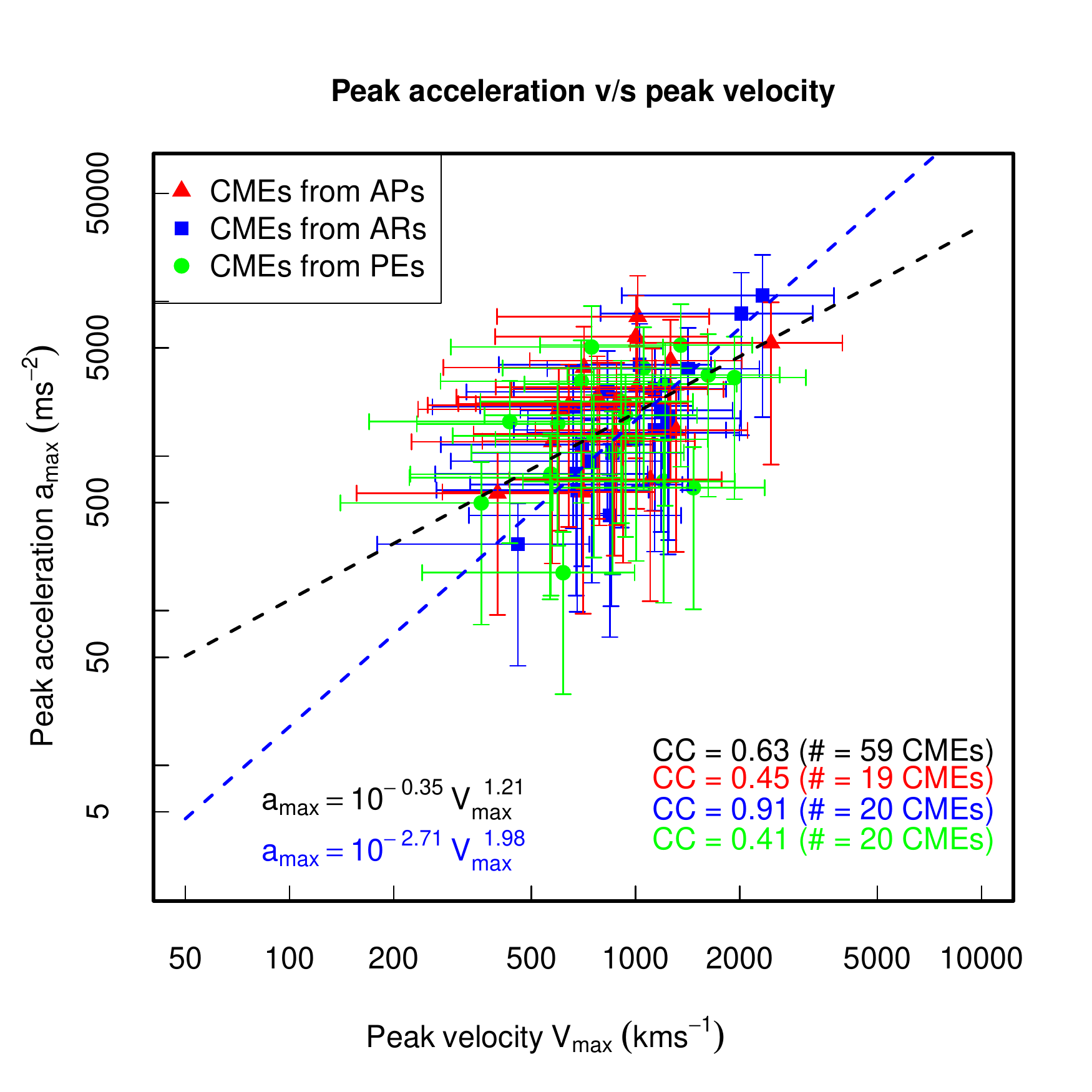}{0.45\textwidth}{(a)}
          \fig{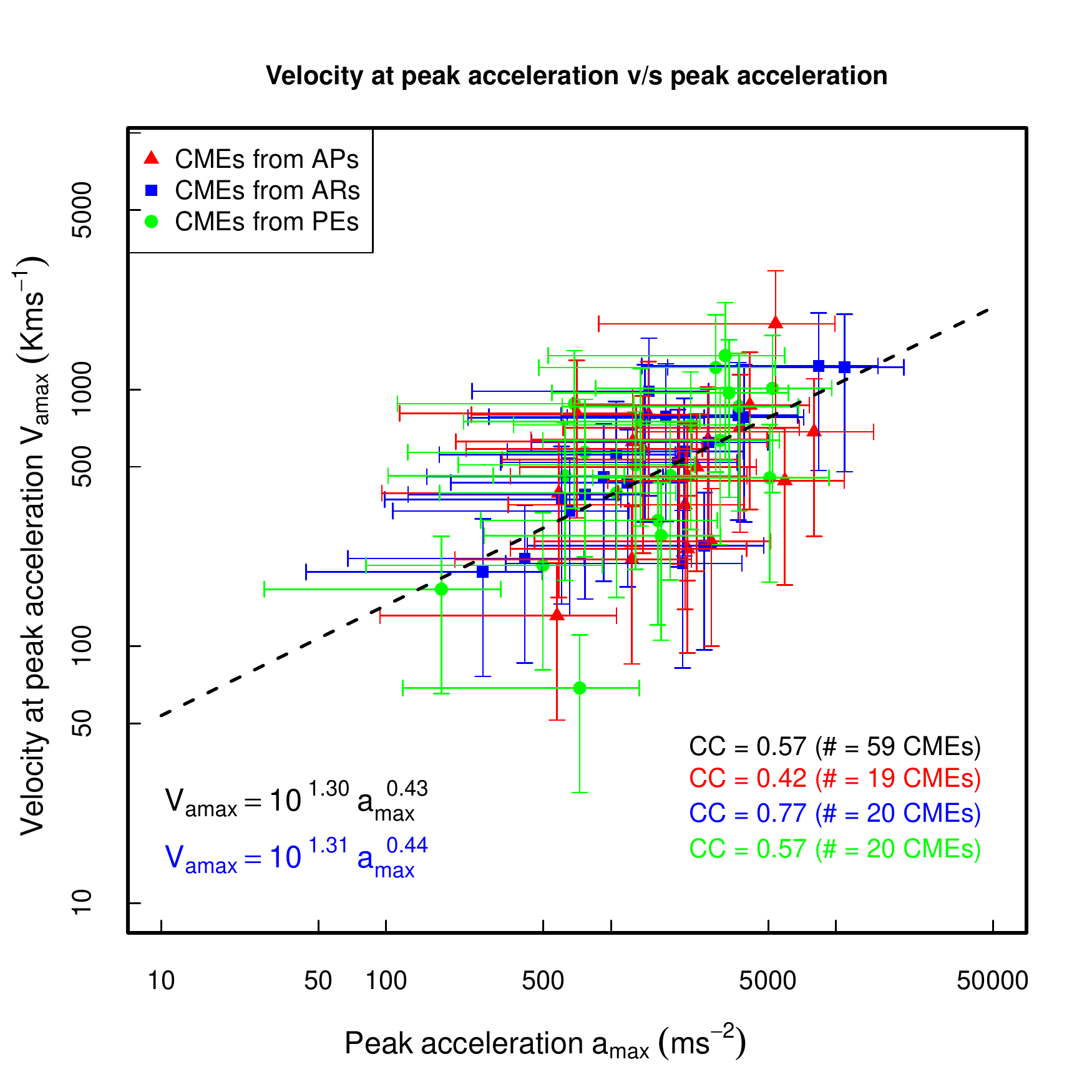}{0.45\textwidth}{(b)}
         }
\caption{ Plot of (a) a$_{\mathrm{max}}$ versus V$_{\mathrm{max}}$ and  (b): V$_{\mathrm{amax}}$ versus a$_{\mathrm{max}}$ of all the CMEs. The dashed curves denote the fitted power-law relation. The data points and fitted curves are color coded according to the source regions.}
\label{fig1}
\end{figure*}

\begin{equation}
   \mathrm{a}_{\mathrm{max}} = 10^{-0.35}\, \mathrm{V}_{\mathrm{max}}^{1.21}.
\end{equation}
A similar result was also reported by \cite{2011ApJ...738..191B} but it must be noted that their numbers suffered from projection effects and their numbers were obtained by combining Extreme Ultraviolet (EUV) and White Light (WL) observations which brings in an additional ambiguity of whether the same physical feature is being tracked in EUV and WL \citep[for a discussion, see][and references therein]{song_2019}, as the former corresponds to the temperature structure of a CME while the later corresponds to the density structure \citep{Ying2020ApJ}. In our work, we do away with both of these limitations, as our measured numbers are in 3D and the measurements are done uniquely in WL.
% , as they found similar power law index with a CC of 0.58 between these quantities. 
Thus, we find that the power law remains unchanged in 3D. Since we also have the information of the source regions of the CMEs, we further find the correlation between these quantities, separately for CMEs coming from different source regions. From this, we find that a$_{\mathrm{max}}$ and V$_{\mathrm{max}}$ are strongly correlated (CC = 0.91) for CMEs coming from ARs, while the ones coming from APs and PEs show weaker correlations of $45\%$ and $41\%$ respectively. This thus indicates at the difference in the CCs for CMEs connected to ARs and CMEs connected to prominences (APs and PEs). We also find that the the two quantities for CMEs from ARs are now related by the relation,
\begin{equation}
    \mathrm{a}_{\mathrm{max}} = 10^{-2.71}\, \mathrm{V}_{\mathrm{max}}^{1.98}.
\end{equation}
Thus, besides arriving at a similar conclusion in 3D as was reported earlier by \cite{2011ApJ...738..191B} in 2D, with the aid of the source region information, we now understand that the source regions have a distinct imprint on the correlation between these parameters, and concluding based on just the overall correlation washes away this crucial information.
For a better understanding, we also plot in Figure \ref{fig1}(b) V$_{\mathrm{amax}}$ versus a$_{\mathrm{max}}$. In this case too we find an overall positive correlation of $0.57$. 
% This tells us that the higher the peak acceleration, the higher is the velocity at that instant. 
This positive correlation can be described by the relation,
\begin{equation}
    \mathrm{V}_{\mathrm{amax}} = 10^{1.30}\, \mathrm{a}_{\mathrm{max}}^{0.43}.
\end{equation}
A similar behaviour was also reported by \cite{Joshi2011ApJ} based on fewer samples, although no such power law relation (or any CC) was reported by them. Further looking into the CCs for CMEs coming from different sources, we find that the ones coming from ARs show a relatively higher correlation with a CC of 0.77 , while it is lesser for those coming from APs and PEs (with CC of 0.42 and 0.57 respectively). Thus once again (in support of our previous result) hinting towards the fact that possibly the dynamics of the CMEs connected to ARs are different from those which are connected to prominences.
% , in a sense that the overall acceleration of the CME is possibly coming from a different mechanism.

\begin{figure*}[h]
\gridline{\fig{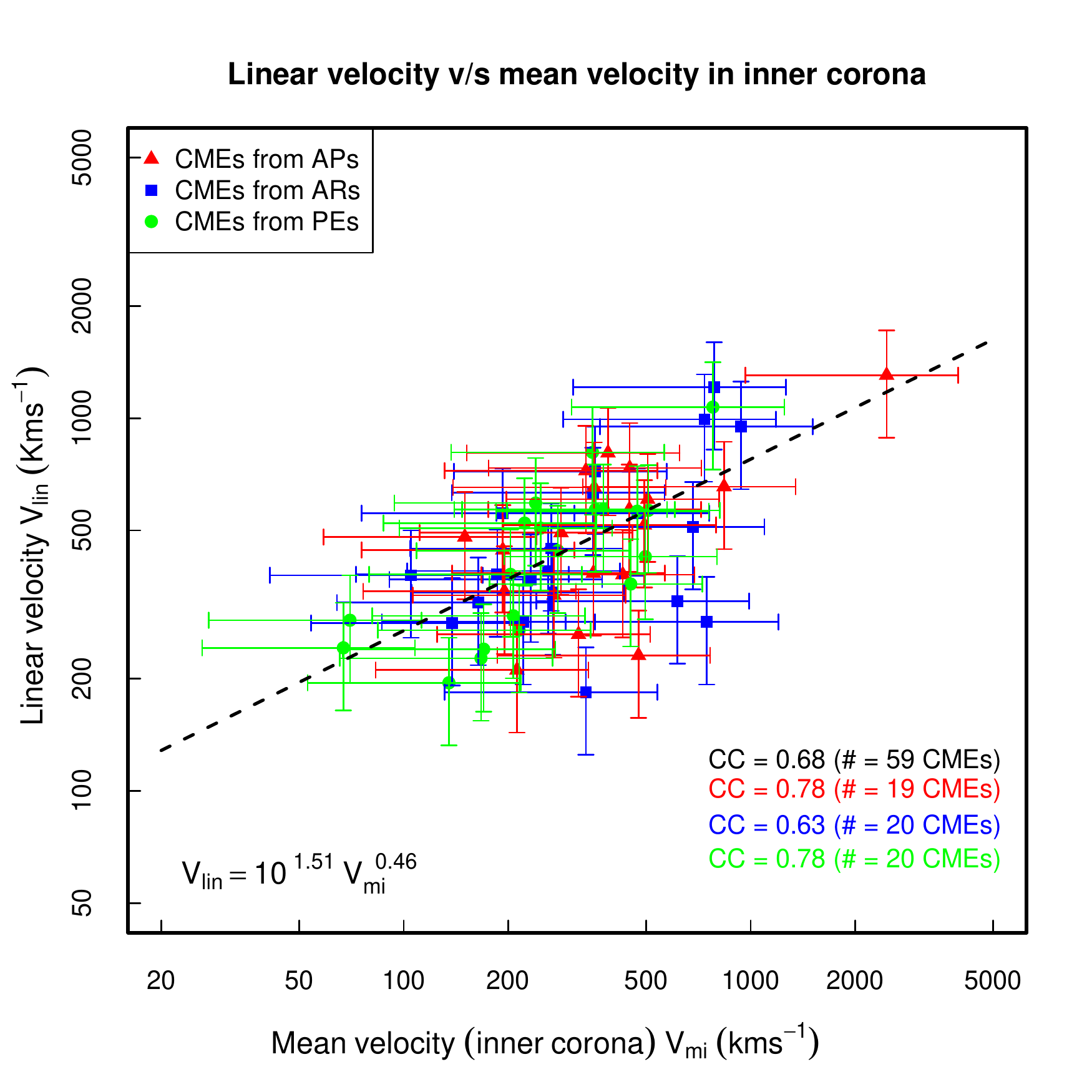}{0.45\textwidth}{(a)}
          \fig{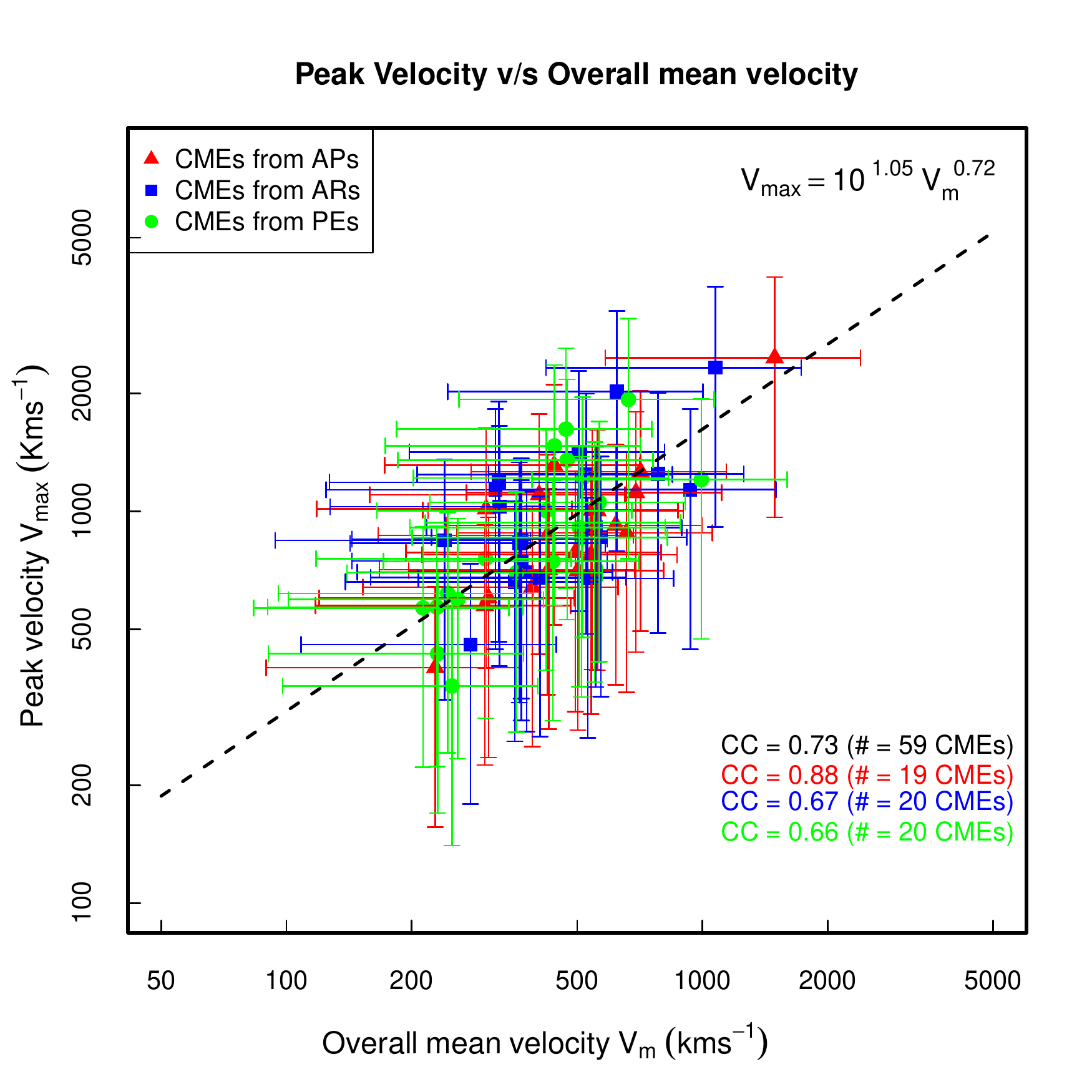}{0.45\textwidth}{(b)}
         }
\caption{Plot of (a) V$_{\mathrm{lin}}$ and V$_{\mathrm{mi}}$ and (b) V$_{\mathrm{max}}$ versus V$_{\mathrm{m}}$. The dashed curves denote the fitted power-law relation. The data points are color coded according to the source regions.}
\label{fig2}
\end{figure*}

Figure \ref{fig2}(a) shows the plot between V$_{\mathrm{lin}}$ and V$_{\mathrm{mi}}$. We find that the two quantities are positively correlated, with an overall CC of 0.68, and they are related by the relation,
\begin{equation}
     \mathrm{V}_{\mathrm{lin}} = 10^{1.51}\, \mathrm{V}_{\mathrm{mi}}^{0.46}. \label{eqn4}
\end{equation}

The CMEs from ARs have a CC of 0.63, while the CMEs connected to prominence eruptions (that is APs or PEs) have a higher correlation of 0.78. This indirectly indicates that such CMEs connected to erupting prominences, experience a small and gradual acceleration that continues during their propagation in the higher heights, while the ones from ARs are more prone to an initial impulsive acceleration followed by a decelerating or constant velocity profile from thereon. Although this was reported in earlier works \citep{Sheeley1999JGR,Moon2002ApJM,article}, it is important to note that this result is based on 3D quantities, and with the relative contribution of different source regions to the overall correlation. Also, from our results we show that similar conclusions (which are well known from previous studies as mentioned above) on the kinematics of gradual and impulsive CMEs can also be obtained from a different perspective by studying the statistical kinematic properties.

Figure~\ref{fig2}(a) also projects the importance of considering the kinematics in the inner corona, and how the kinematic parameters in the outer corona is influenced by those in the inner corona. Also, several models of CME arrival time predictions, use V$_{\mathrm{lin}}$ as input to calculate their arrival times  \citep[for a review see][]{2014SpWea..12..448Z,2018SpWea..16.1245R}. An important aspect in this regard is the lead time of forecast, which is the difference between the observed arrival time of CME, and the time of submission of the forecast \citep{2018SpWea..16.1245R}. With the help of equation \ref{eqn4}, the lead time can be minimised, by estimating V$_{\mathrm{lin}}$ from V$_{\mathrm{mi}}$, thereby reducing the absolute dependency on outer corona observations for arrival time predictions. Please note that an estimate of the gain in lead time would be possible only with the availability of near-real time data. Apart from that, it will also depend on the telemetry rate of the instruments involved, as that would dictate the accessibility of the near real time data. Having said that, we would also like to emphasize that our result will help in the implementation of inner coronal observations from space missions like ADITYA-L1 \citep[][]{aditya-l1,prasad_velc_2017} and PROBA-3 \citep[][]{proba-3} (which do not have outer coronal observations) solely for the purpose of arrival time estimation of CMEs.  

In Figure \ref{fig2}(b), we plot V$_{\mathrm{max}}$ versus V$_{\mathrm{m}}$ and find that the two quantities are positively correlated, with an overall CC of 0.73, and they can be related by the relation,
\begin{equation}
     \mathrm{V}_{\mathrm{max}} = 10^{1.05}\, \mathrm{V}_{\mathrm{m}}^{0.72}.
     \label{eqn5}
\end{equation}

This throws light on the kind of acceleration profiles, the CMEs experienced. The ones experiencing impulsive acceleration also experience a high retardation \citep[see Figure 2 in][]{2020ApJ...899....6M} which will largely affect V$_m$, while the ones experiencing uniform acceleration do not experience such high retardation. Again, looking at the source region contributions, we see that the ones from ARs have a comparatively lower CC of 0.67, and similar value for the ones from PEs, while the CMEs from APs shows a higher CC of 0.88. This thus re-establishes that CMEs connected to ARs are mostly impulsive ones and experience higher deceleration than those connected to APs, which is leading to a lower correlation in the former. A lower correlation in the case of CMEs from PEs again point that they are gradual events where a steady small acceleration prevents the mean velocity to correlate more strongly with the peak velocity (as the later keeps on increasing). It is important to note that this was also pointed out by \cite{2020ApJ...899....6M}, and thus we re-affirm those results in a more statistical manner here. The above relation (Equation~\ref{eqn5}) could also be empirically used to estimate one of the quantities when the other is known. For example, if we have inner corona observations in the future from Aditya-L1/VELC and STEREO/COR-1A, then V$_{\mathrm{max}}$ could be measured in inner corona from which the mean speed (V$_{\mathrm{m}}$) could be estimated based on such empirical relation quickly. V$_{\mathrm{m}}$ has further different applications such as in drag based models for CME propagation.

\begin{figure*}[h]
\gridline{%\fig{amax_vin.pdf}{0.45\textwidth}{(a)}
          \fig{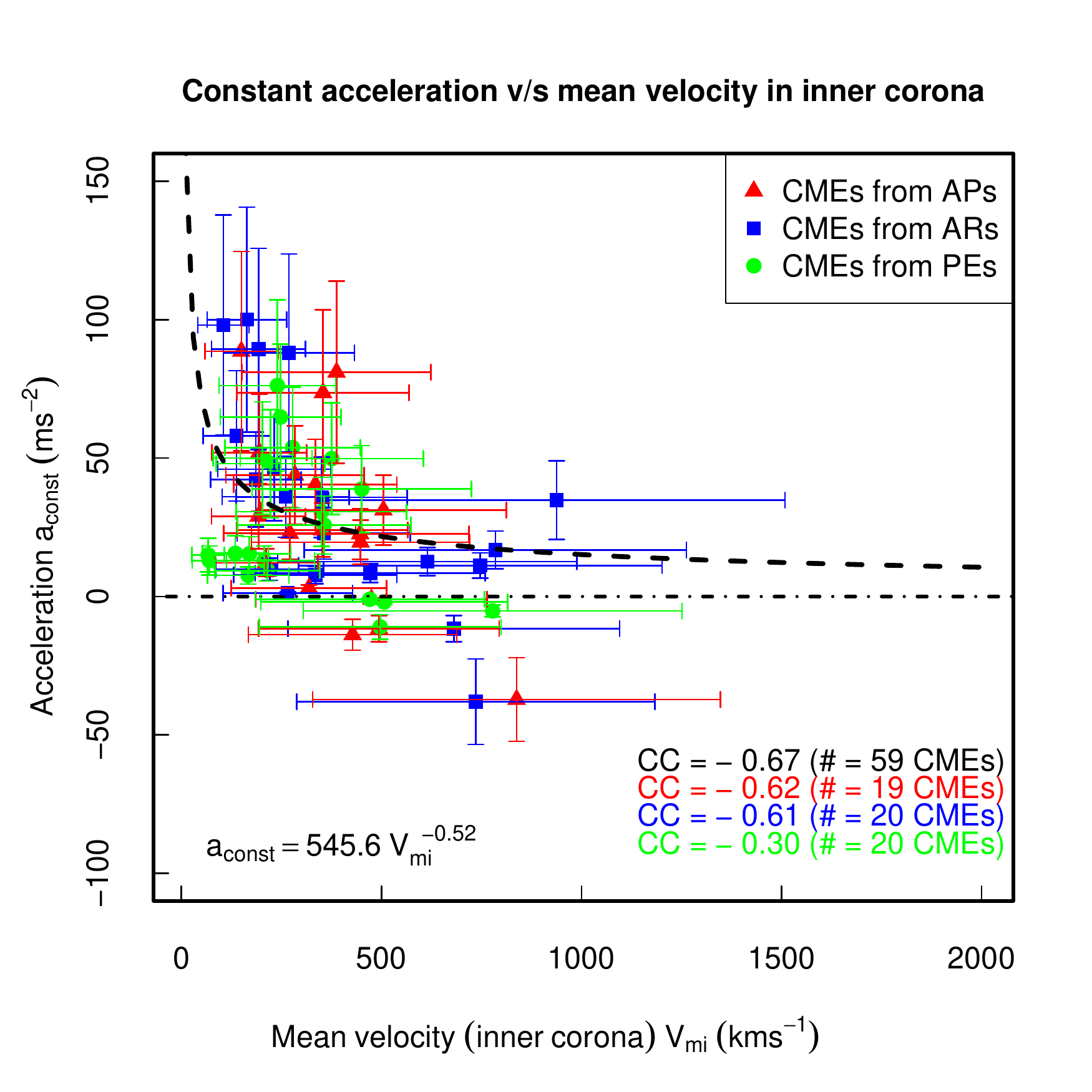}{0.45\textwidth}{(a)}
         }
\caption{Plot of (a) a$_{\mathrm{const}}$ versus V$_{\mathrm{mi}}$. The dashed curves denote the fitted power-law relation. The data points are color coded according to the source regions.}
\label{fig3}
\end{figure*}
We look into the correlation between a$_{\mathrm{const}}$ and V$_{\mathrm{mi}}$ in Figure \ref{fig3}(a). We find a clear anti-correlation between the two quantities with a CC of -0.67, where the two quantities are related by

\begin{equation}
    \mathrm{a}_{\mathrm{const}} = 545.6\, \mathrm{V}_{\mathrm{mi}}^{-0.52}.
\end{equation}

This indicates at the interaction of the CME and the solar wind and hence the drag experienced by the former due to the later. It is worthwhile to note that such acceleration velocity anti-correlation has been reported earlier \citep[see][]{Moon2002ApJM,vrsnak2004A&A}, but such reports only include results in the outer corona and with projected values of acceleration and velocities (starting with projected height of at least 2 R$_\odot$). In our work, we report on a similar anti-correlation that exists in the inner corona as well, and our result is based on 3D acceleration and velocity. This anti-correlation thus shows that the influence of the drag forces comes into play as early as in the inner corona.  Further, on looking into the individual source region contributions, we find that the CMEs from ARs have a CC of -0.61, and a similar correlation (-0.62) for CMEs from APs, while this anti-correlation is relatively poor for PEs (-0.30). This distinct difference in the CC for CMEs from PEs and CMEs from active regions (ARs and APs) points to the contrasting acceleration experienced by the CMEs. CMEs with impulsive acceleration are faster, which increases the drag, and hence a higher retardation which is reflected in the higher value of anti-correlation in the case of CMEs connected to ARs. This is not the case for CMEs from PEs which are predominantly gradual events, and hence a weaker anti-correlation reflecting a weaker drag experienced. Thus we again find the importance of the source region information in the study of statistical kinematics of CMEs.

\begin{figure*}[h]
\gridline{\fig{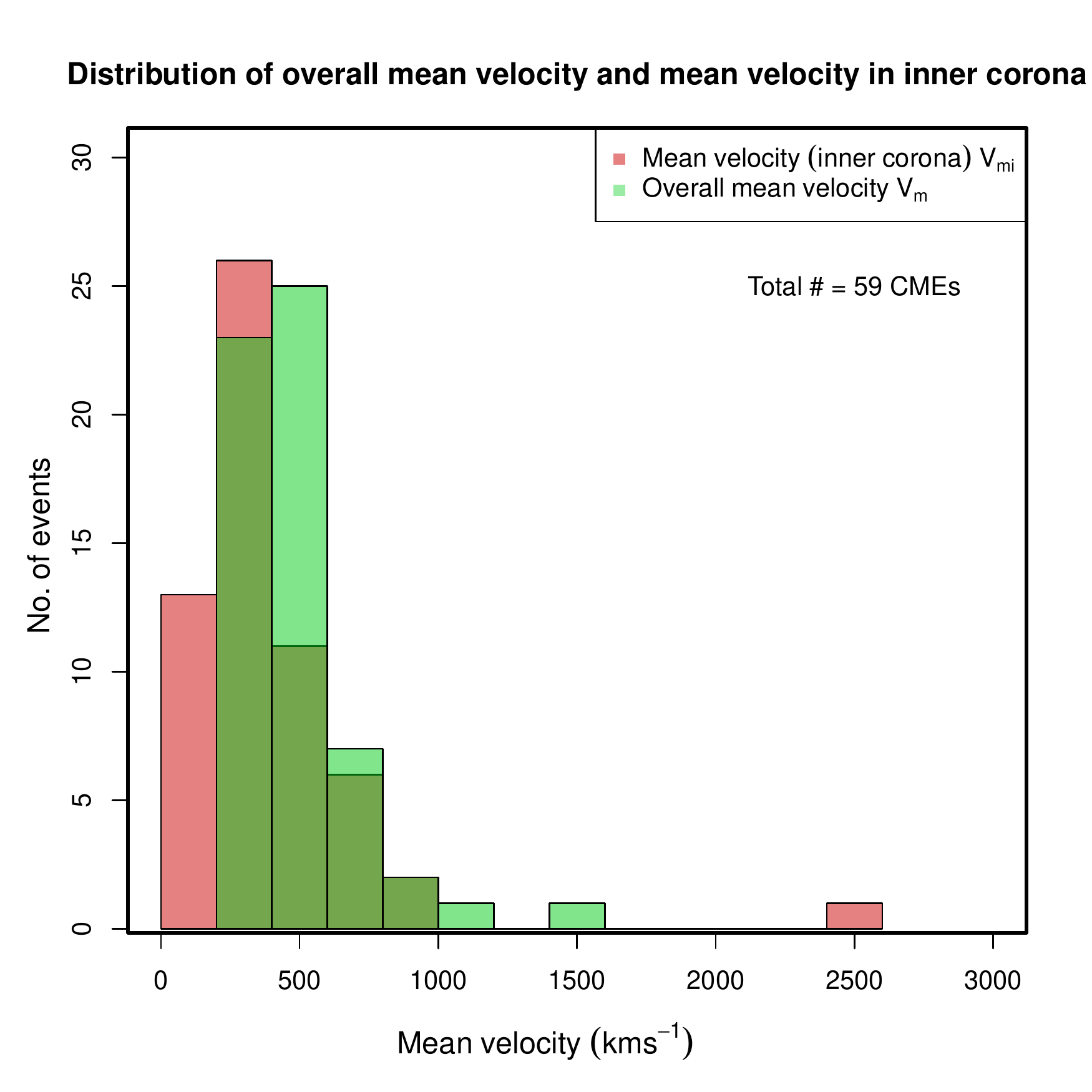}{0.45\textwidth}{(a)}
          \fig{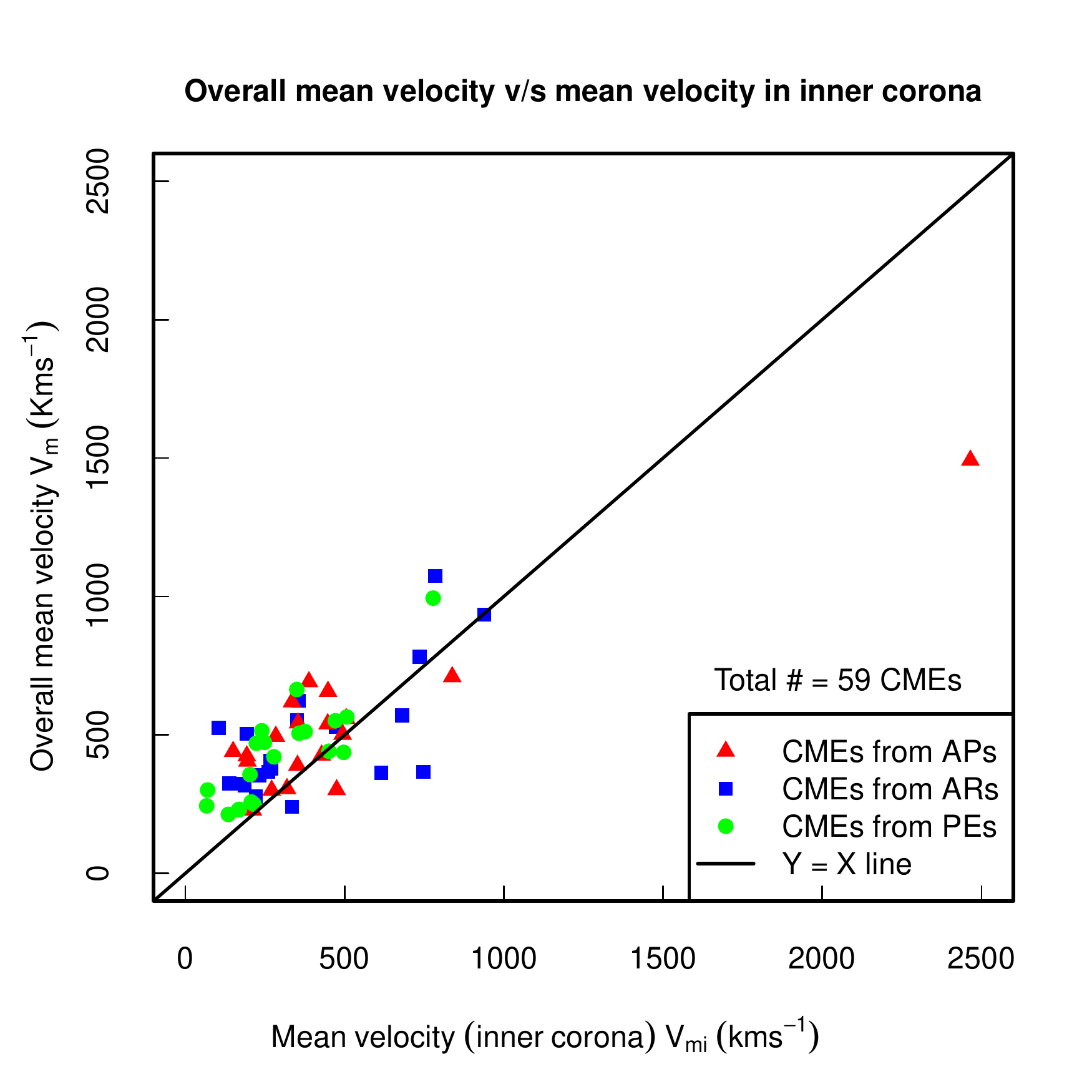}{0.45\textwidth}{(b)}
         }
\gridline{\fig{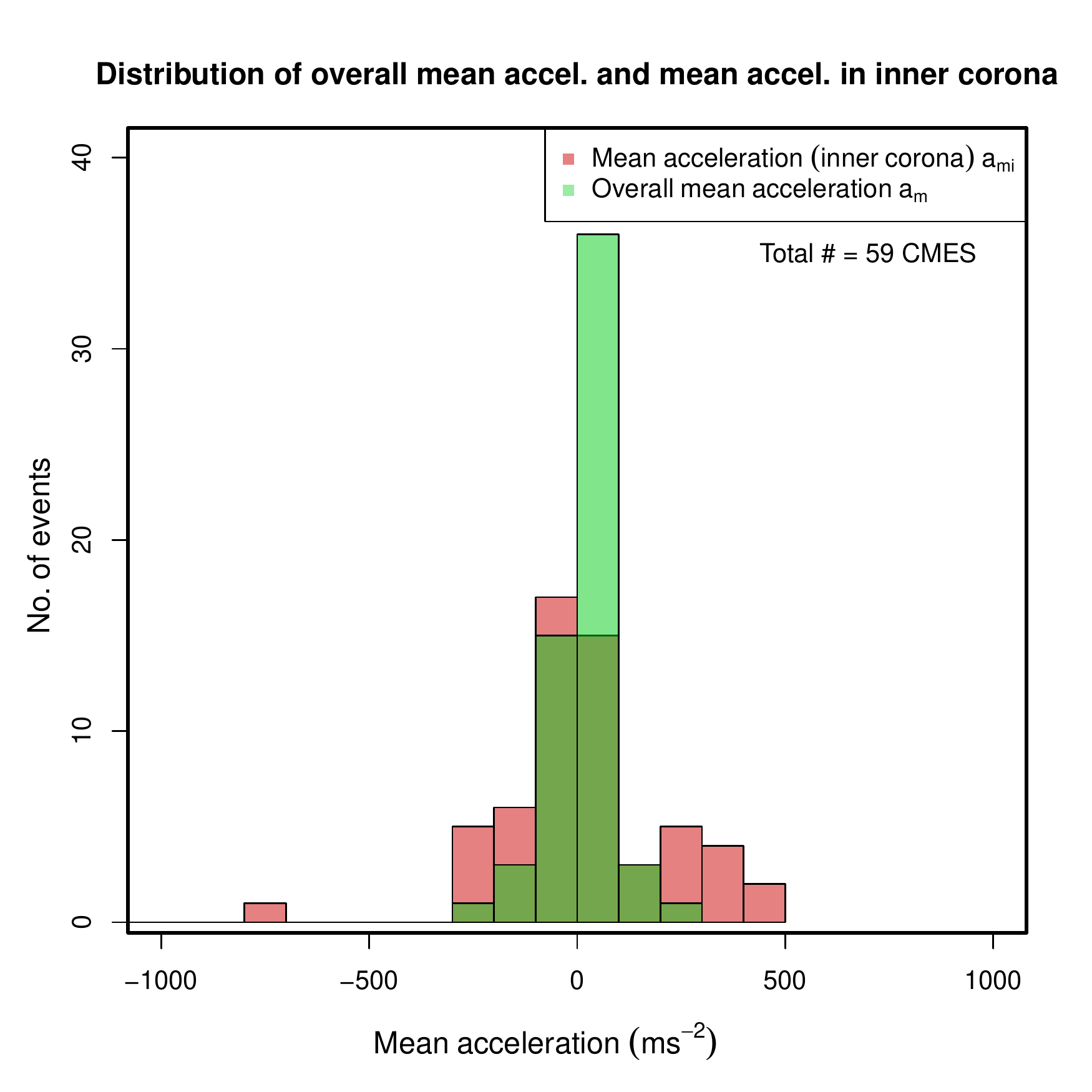}{0.45\textwidth}{(c)}
         \fig{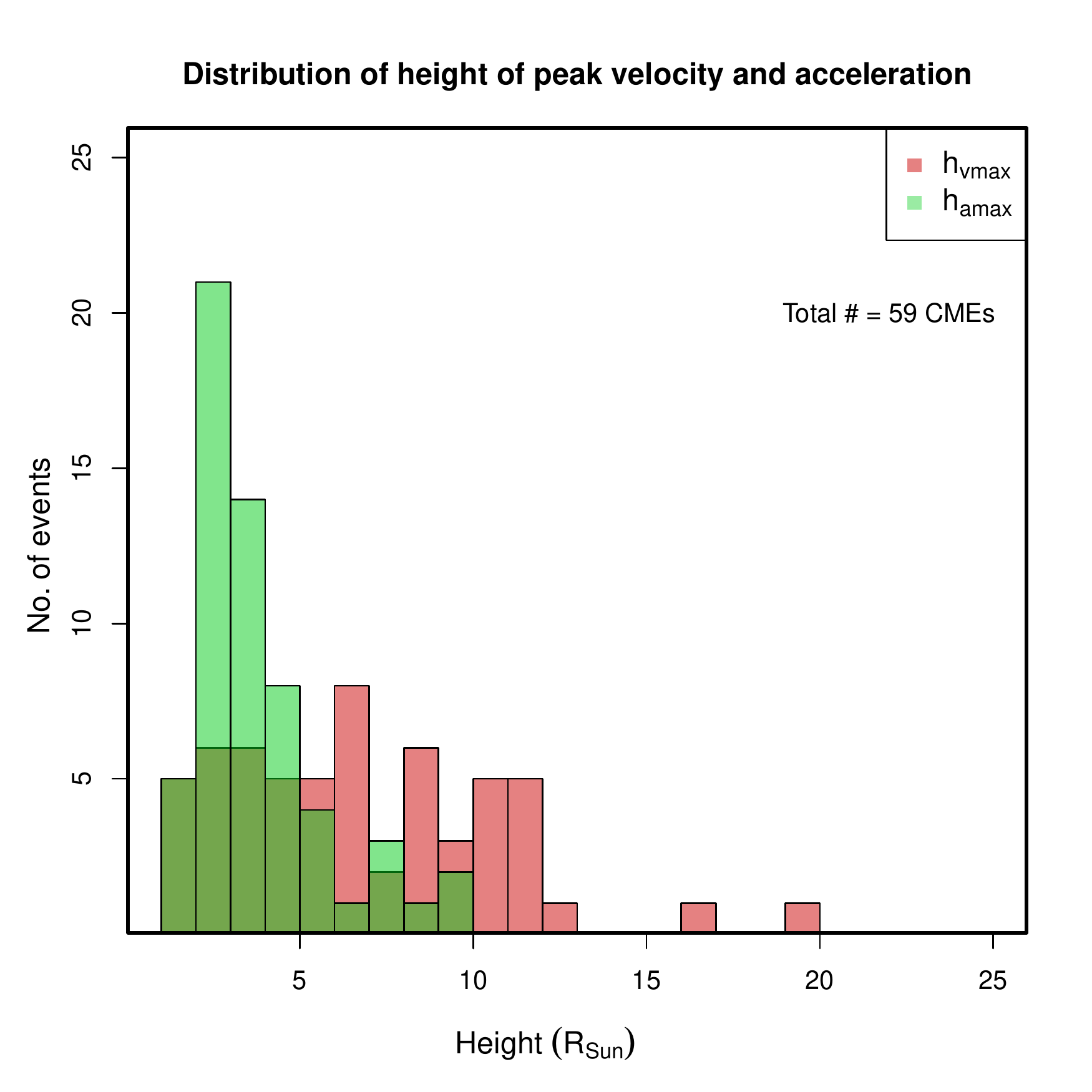}{0.45\textwidth}{(d)}
         }
\caption{(a): Distribution of V$_{\mathrm{mi}}$ and V$_\mathrm{m}$ (b): V$_{\mathrm{mi}}$ versus V$_\mathrm{m}$. (c) Distribution of a$_{\mathrm{mi}}$ and a$_\mathrm{m}$. (d) Distribution of h$_{\mathrm{amax}}$ and h$_{\mathrm{vmax}}$.The data points are color coded according to the source regions.}
\label{fig4}
\end{figure*}

In Figure~\ref{fig4}(a) we plot the distribution of V$_{\mathrm{mi}}$ and V$_\mathrm{m}$. We find that the distribution of V$_\mathrm{m}$ has been shifted towards the right side with respect to the distribution of V$_{\mathrm{mi}}$. Thus, for average quantities, it is important to specify the region where the average has been taken, as we can see that the numbers change from the inner to the outer corona. For a better illustration, in Figure \ref{fig4}(b) we plot V$_\mathrm{m}$ versus V$_{\mathrm{mi}}$. The solid line represents the boundary where both the quantities are equal. We see that for most of the CMEs, V$_\mathrm{m}$ is greater than V$_{\mathrm{mi}}$ implying that the CMEs have gained speed while propagating in the outer corona. We also note that for CMEs which have V$_{\mathrm{mi}}$ greater than V$_\mathrm{m}$, most of them are from ARs, thus re-indicating the presence of impulsive accelerations in the lower heights, followed by deceleration later. Also, the CMEs coming from PEs have V$_\mathrm{m}$ greater than V$_{\mathrm{mi}}$, thus confirming that they experience gradual acceleration for a longer duration that lead to a steady increase in their velocities as they propagate outwards. So, we note that working with a single average velocity of the CME for its entire trajectory, often masks this important information. Similarly, in Figure \ref{fig4}(c) we plot the distribution of a$_{\mathrm{mi}}$ and a$_\mathrm{m}$. We find that the distribution of a$_{\mathrm{mi}}$ was relatively more spread out around the zero value, thus indicating acceleration and deceleration and hence the wide range of kinematics exhibited by CMEs connected to different source regions, while the distribution of a$_\mathrm{m}$ is more narrowed around the zero value with the mode of the distribution lying in the range of 0-100 m s$^{-2}$. Thus showing that the CMEs experience very little acceleration in their higher heights. For a better understanding, in Figure \ref{fig4}(d) we plot the distribution of h$_{\mathrm{amax}}$ and h$_{\mathrm{vmax}}$. In support to our former argument on Figure \ref{fig4}(c), here too we see that the distribution of h$_{\mathrm{amax}}$ is not as spread out as the distribution of h$_{\mathrm{vmax}}$. We also see that the mode of the distribution of h$_{\mathrm{amax}}$ lies at 2-3 R$_{\odot}$, which is also supported by the results of \cite{2020ApJ...899....6M} on the fact that the impact of Lorentz force on the 3D kinematics of CMEs stays dominant till a height range of 2.5-3 R$_{\odot}$. Thus, this result also points to the  fact that it is also the Lorentz force that is closely responsible for the CMEs attaining their peak accelerations during their propagation. While looking into the distribution of h$_{\mathrm{vmax}}$, we find that we do not get to see any such clear peak in the distribution, and that it is much more spread out. This is possibly because we have selected events from three different classes, CMEs from ARs which show presence of impulsive acceleration which occur for a short duration and CMEs from PEs which show presence of small gradual accelerations that occur for a longer duration. 

%% The "ht!" tells LaTeX to put the figure "here" first, at the "top" next
%% and to override the normal way of calculating a float position

\section{Summary and Conclusions} \label{conclusion}
We have studied the behaviour of several 3D kinematic parameters of the 59 CMEs studied by \cite{2020ApJ...899....6M} and we extended their analysis in this work. Several correlations studied between different kinematic parameters showed the importance of considering inner corona observations in the understanding of CME kinematics, and how different kinematic parameters in the outer corona are influenced by the parameters in the inner corona. We also found that the overall correlations often washes away crucial information and individual correlations for CMEs from different source regions show the imprint of source regions on the kinematics. In this regard, the change in the power law exponent for the different CCs is not much pronounced which has led to a considerable overlapping of the data points for different source regions. Recently, \cite{pant_2021} reported on the clear influence of the source region on the width distribution of slow and fast CMEs, concluding on the possibility of different physical ejection mechanisms for the CMEs from ARs and PEs. In this work, we thus look into the correlations of several kinematic parameters, and we again find a clear imprint of the source regions (in the form of distinctly different individual CCs) on the overall correlations. Further, we find that while working with average kinematic quantities, it is important to specify the region where the average is taken, as the average values change with the CME propagating from the inner to the outer corona. It should also be noted that even within the inner corona, the average values change, and in this work, we pointed out this, that tagging a single average speed to a CME might not be the best way to comment on its speed and hence we show as an example that the average speed of a CME indeed changes as the CME travels from the inner to the outer corona. It is also worth noting that our results are based on 3D parameters and are hence independent of projection effects. In the following, we conclude our main results,

\begin{itemize}
\item A study of a$_{\mathrm{max}}$ and V$_{\mathrm{max}}$ revealed that the two quantities are positively correlated with a CC of 0.63. The CC is significantly higher (0.91) for CMEs from ARs as compared to the ones from APs and PEs. A study of a$_{\mathrm{max}}$ and V$_{\mathrm{amax}}$ further showed that despite a moderate overall correlation, the ones coming from ARs show a much higher positive correlation (CC 0.77), indicating that the maximum velocity and accelerations are better correlated for CMEs from ARs.
    
\item We found V$_{\mathrm{lin}}$ and v$_{\mathrm{mi}}$ to be positively correlated (CC 0.68), and that the former can be estimated from the later through equation \ref{eqn4}, thus enabling the use of inner coronal observations to CME arrival time predictions with better lead time of forecast. Further a study of V$_{\mathrm{max}}$ and v$_\mathrm{m}$ indicated indirectly at the acceleration experienced by the CMEs from different source regions. 
    
\item We also found an anti-correlation between a$_{\mathrm{const}}$ and V$_{\mathrm{mi}}$ with a CC of -0.67 that shows evidence of the drag experienced by the CME due to interaction with the solar wind. Thus showing that the influence of the drag forces comes into play as early as in the inner corona. While, the CMEs from ARs and APs have similar CCs, for PEs, the correlation is much weaker with a CC of -0.30.

\item From Figure \ref{fig4}(a) and (b), we found the average velocities change as the CMEs travel from the inner to the outer corona, and that the CMEs from PEs experience weak and gradual accelerations, while the ones from ARs experience impulsive accelerations followed by retardation. This was further supported by Figure \ref{fig4}(c) which showed that a$_\mathrm{m}$ is more confined around the zero value while a$_{\mathrm{mi}}$ is relatively more spread about the zero value. Also, we found the distribution of h$_{\mathrm{amax}}$ peaks around $2-3$ R$_\odot$, which supports the results of \cite{2020ApJ...899....6M} on the fact that the impact of Lorentz force stays dominant in 2.5-3 R$_{\odot}$. Thus indicating the role of Lorentz force in propelling the CMEs to their peak accelerations.
\end{itemize}

A number of upcoming space missions like ADITYA-L1, PROBA-3 and the recently launched Solar Orbiter \citep{2013SoPh..285...25M} will be observing the inner corona, and we believe these results will provide rich inputs to their observation plans. Also, the above correlations correlating the parameters in the inner corona to the parameters in the outer corona will help in better exploiting the inner coronal observations. Further extending this study over a larger sample size, will help in better establishing our claims. The results will also provide inputs to models that study CME ejection mechanisms, thus aiding in our present understanding of the same.

We thank the anonymous referee for the valuable comments that have improved the manuscript. We would also like to thank IIA for providing the necessary computational facilities for this work. The SECCHI data used here were produced by an international consortium of the Naval Research Laboratory (USA), Lockheed Martin Solar and Astrophysics Lab (USA), NASA Goddard Space Flight Center (USA), Rutherford Appleton Laboratory (UK), University of Birmingham (UK), Max-Planck-Institut for Solar System Research (Germany), Centre Spatiale de Li$\grave{e}$ge (Belgium), Institut d'Optique Th$\acute{e}$orique et Appliqu$\acute{e}$e (France), Institut d'Astrophysique Spatiale (France). The CDAW CME catalog is generated and maintained at the CDAW Data Center by NASA and The Catholic University of America in cooperation with the Naval Research Laboratory. We also acknowledge SDO team for making AIA data available and SOHO team for EIT and LASCO data. SOHO is a project of international cooperation between ESA and NASA.

%% Putting eqnarrays or equations inside the mathletters environment groups
%% the enclosed equations by letter. For instance, the eqnarray below, instead
%% of being numbered, say, (4) and (5), would be numbered (4a) and (4b).
%% LaTeX the paper and look at the output to see the results.

\acknowledgments


\vspace{5mm}

\end{document}